\documentclass[final,3p,times]{elsarticle}

\usepackage{amssymb}
\usepackage{amsmath}
\usepackage{natbib}
\usepackage{subcaption}
\usepackage{hyperref}
\usepackage{cleveref}
\hypersetup{
    colorlinks,
    linkcolor={red},
    citecolor={blue},
    urlcolor={blue},
    urlbordercolor=1 0 0
}

\begin{document}

\begin{frontmatter}



\title{Tackling inverse problems for PDFs from lattice QCD}


\author{Alexander Rothkopf} 
\ead{akrothkopf@korea.ac.kr}

\affiliation{organization={Department of Physics, Korea University},
            city={Seoul},
            postcode={02841}, 
            country={Republic of Korea}}

\begin{abstract}
In this kick-off presentation for the "Recent developments in QCD" session at Baryons 2025 I will tie together the recent progress made on the extraction of parton distribution functions (PDFs) in lattice QCD and the long standing efforts in solving the inverse problem in the form of spectral function reconstruction. 
\end{abstract}

\begin{keyword}
Parton Distribution Functions \sep Lattice QCD \sep Inverse Problem
\end{keyword}

\end{frontmatter}



\section{Parton Distribution Functions}

\subsection{Introduction}
Parton distributions functions (PDF) are a key theory tool in the study of hadron structure and lepton-nucleus scattering \cite{Amoroso:2022eow}. They are indispensable in the interpretation of the production of e.g. hard probes in p+p, as well as Pb+Pb collisions at the Large Hadron Collider at CERN but will form a central research focus on their own right at the upcoming electron-ion collider at Brookhaven National Laboratory \cite{eicphi2026phenomenological}. A precision understanding of neutrino-nucleus scattering \cite{Formaggio:2012cpf} on the other hand is vital for large scale experiments searching for physics beyond the standard model and contingent on detailed knowledge of PDFs.

Parton distribution functions are defined with deep inelastic scattering in mind (for an accessible introduction see \cite{soper1997parton}), in which a lepton $\ell$ scatters off a nucleon by exchanging a virtual photon. In fact, the virtual photon will interact with one of the partons within the nucleon, which travel close to the speed of light. It has been shown (for a review see e.g. \cite{collins1987theorems}) that the cross-section of such a scattering process can be {\it factorized} into a perturbative partonic cross-section and the non-perturbative PDF. The latter encodes the probability to find a parton $a$ with a momentum fraction $x_a$ (Bjorken-x) inside the nucleon $n$
\begin{align}
d\sigma_{\ell n} \sim \sum_a \int dx_a f_a(x_a,\mu) d\sigma_{\ell a} . \label{eq:PDF}
\end{align}

In order to describe the motion of the partons close to the lightcone one switches to the eponymous coordinate system, which for a nucleon beam along the z-axis reads
\begin{align}
x^{\pm} = (x^0\pm x^3)/\sqrt{2}, \quad P^{\pm} = (P^0\pm P^3)/\sqrt{2}, \quad P\cdot x = P^+x^- + P^-x^+ - {\bf P}_T\cdot {\bf x}_T \label{eq:lightconecord}.
\end{align}

Since in the scattering process of interest, the lepton couples to the parton via a gauge vector field, the correlation function we need to evaluate is that of a point-split vector current evaluated on nucleon states with momentum $P^+$ along the lightcone. As neither color nor spin structure is essential for the subsequent discussion of kinematic properties, let us follow Ref.~\cite{radyushkin2017quasi} and strip away the technical intricacies of QCD and consider simply the matrix element of a bilocal operator evaluated in a momentum eigenstate
\begin{align}
\langle p | \phi(0) \phi(z)|p\rangle = {\cal M}(-(pz), -z^2).
\end{align}
Due to Lorentz invariance ${\cal M}$ can only depend on Lorentz invariant combinations of $p$ and $z$ one of which, $-pz=\nu$ is christened Ioffe time.

If one considers the Fourier transform in the Ioffe time argument, 
\begin{align}
 {\cal M}(\nu, -z^2)=\int_{-1}^1 dx \, e^{ i \nu x} {\cal P}(x,-z),
\end{align}
one can show that the conjugate Fourier variable $-1\leq x \leq1$ has compact support and offers a covariant definition of the intuitive concept of momentum fraction carried by a parton. The physical PDF $f(x)$ is related to the matrix element ${\cal P}$ evaluated on the light cone. In order for the scattering process to take place the interaction vertex must annihilate a field with momentum $(xP^+)$, which is possible if $z$ is chosen to be light-like in the minus direction (see \cref{eq:lightconecord}). In that case
\begin{align}
{\cal M}(-P_+ z_-, 0)=\int_{-1}^1 dx \, f(x)\,  e^{ - i P_+ z_-}, \qquad f(x) =\frac{1}{2\pi}\int_{-\infty}^{\infty} d\nu \, e^{-ix\nu}\, {\cal M}(\nu,0) = {\cal P}(x,0).
\end{align}
We find that we can recover the actual PDF $f(x)$ from the $-z^2\to 0$ limit of the matrix element {\cal P}. 

When we introduced PDFs in \cref{eq:PDF} they carried a scale dependence $\mu$ which appears to be absent here, but which will be introduced by the renormalization procedure. It turns out that the matrix element exhibits running with ${\rm log}(z^2)$, indicating that a probe with a higher resolving power (associated with the energy scale $\mu$) will see more virtual processes involving quarks temporarily emitting and reabsorbing gluons.

\subsection{Two effective approaches to PDFs on the lattice}

If we wish to evaluate PDFs on the lattice, we face the fundamental limitation that lattice QCD simulations are carried out in Euclidean time and only access to space-like momenta is possible. A direct evaluation of the matrix element {\cal P} on the lightcone is therefore out of the question. Over the past decade very fruitful ideas emerged, which suggest to approach the lightcone from the space-like region in order to systematically approximate the sought after real-time physics from a Euclidean lattice. It is interesting to note that the idea of Euclideanization of light-cone physics \cite{caron2009g} has been independently explored in connection with in-medium jet quenching \cite{Panero:2013pla} around the same time. 

In the \textit{quasi PDF} approach, pioneered in Ref.~\cite{ji2013parton} one may choose a space-like four vector, e.g. $z=(0,0,0,z_3)$, leading to an equal-time correlation function 
\begin{align}
\langle p | \phi(0) \phi(z)|p\rangle = {\cal M}(\nu,z_3^2)=\int_{-\infty}^\infty dy\, Q(y,P) \, e^{iy P z_3}.
\end{align}
The quasi PDF $Q(y,P)$ may be obtained from the inverse Fourier transform (IFT) over Ioffe time $\nu$
\begin{align}
Q(y,P) = \frac{1}{2\pi} \int_{-\infty}^\infty d\nu e^{-i y \nu} {\cal M}(\nu,\nu^2/P^2).
\end{align}
From the above relation it is clear that the quasi PDF agrees with the physical PDF in the limit of large momentum $P$. In this approach the inverse Fourier transform is performed to obtain $Q$ and only subsequently renormalization of the PDF is carried out.

An alternative procedure is the \textit{pseudo PDF} approach introduced in Ref.~\cite{radyushkin2017quasi}. Here the matrix element ${\cal M}(\nu,z^2)$ is first cleaned of UV divergences by forming the ratio $\mathbb{M}(\nu,z^2)={\cal M} (\nu,z^2)/{\cal M} (0,z^2)$. This reduced matrix element can in turn be related to the renormalized pseudo PDF $Q_{\bar{MS}}(\nu,\mu^2)$ via matching. Through matching, the dependence on $z$ will be converted into a dependence on the matching (renormalization) scale $\mu$. It is now, after renormalization, that the necessary inverse Fourier transform is performed
\begin{align}
Q_{\bar{MS}}(\nu,\mu^2)=\int_{0}^1 dx \cos(\nu x) f(x,\mu^2)\label{eq:defpseudo},
\end{align}
to extract the physical PDF $f(x,\mu)$ from the Ioffe time pseudo PDF.

Whereas the need for evaluating the matrix element is explicit in the quasi PDF approach it also enters in the pseudo PDF approach as the reach within Ioffe time $\nu=-Pz_3$ depends on the accessible values of momenta.

In order to assess the challenge associated with the large momentum limit in the quasi PDF approach, Ref.~\cite{nam2017quasi} explored its convergence properties for pion quasi distribution amplitudes in a nonlocal chiral-quark model. It was found that in order to recover the fourth moment of the physical pion distribution amplitude to percent accuracy, access to momenta of around $30$GeV is necessary.

For other approaches to PDFs on the lattice see see also the work on the hadronic tensor (see e.g. Ref.~\cite{liu2015pos}), Compton amplitudes (see e.g. \cite{Can:2022chd}) and good lattice cross sections (see e.g. \cite{Ma:2014jla}). For reviews see e.g. Refs.~\cite{ji2021large,constantinou2021x}.

\section{Defining and spotting the inverse problem}

In the preceding section we found that the inverse Fourier transform constitutes a key ingredient in the extraction of PDFs from matrix elements, be it before or after renormalization. At first sight it may appear that such a transformation is well-posed. This however is contingent on our access to the Brillouin zone. Let us take a closer look at the pseudo PDF setting of \cref{eq:defpseudo}. On the lattice we are dealing with discrete momenta and thus discrete values of Ioffe time $\nu_k$ such that $\mathfrak{Q}_k =Q_{\bar{MS}}(\nu_k,\mu^2)$. For numerical treatment we discretize the PDF ${\mathfrak f}_j=f(x_j,\mu^2)$ and introduce a quadrature of the integral over $N$ discrete $x$ values, turning the convolution into a vector-matrix multiplication $\mathfrak{Q} = \mathfrak{K}\cdot {\mathfrak f}$. Choosing the trapezoid rule we can write the kernel matrix as $\mathfrak{K}_{ij}=\kappa \frac{1}{N} \cos(\nu_k x_l)$ where $\kappa=\frac{1}{2}$ for $l=1$ or $N$ and $\kappa=1$ otherwise.

\begin{figure}
	\begin{subfigure}[t]{0.33\textwidth}
		\includegraphics[scale=0.35]{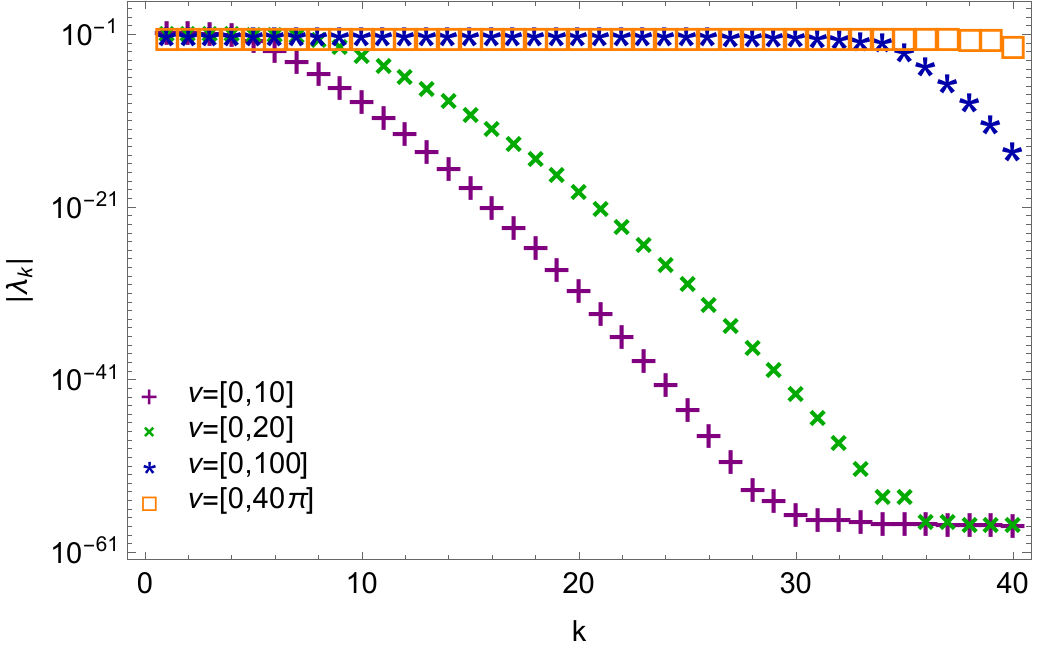}
		\caption{}
	\end{subfigure}\hspace{1cm}
	\begin{subfigure}{0.60\textwidth}
		\includegraphics[scale=0.3]{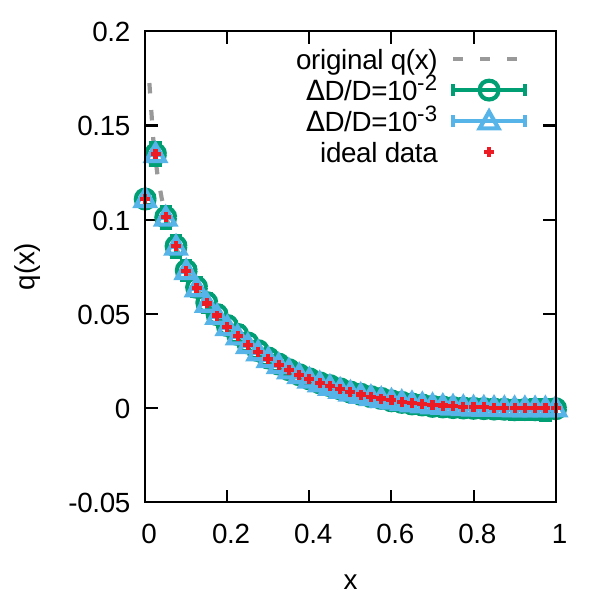}
		\includegraphics[scale=0.3]{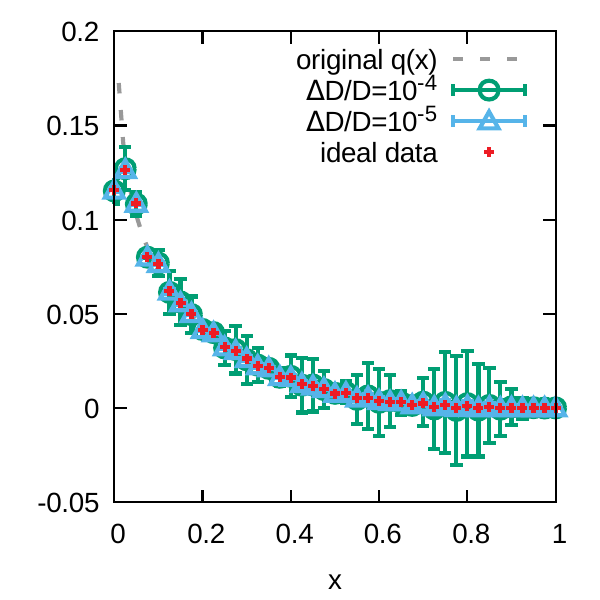}
		\includegraphics[scale=0.3]{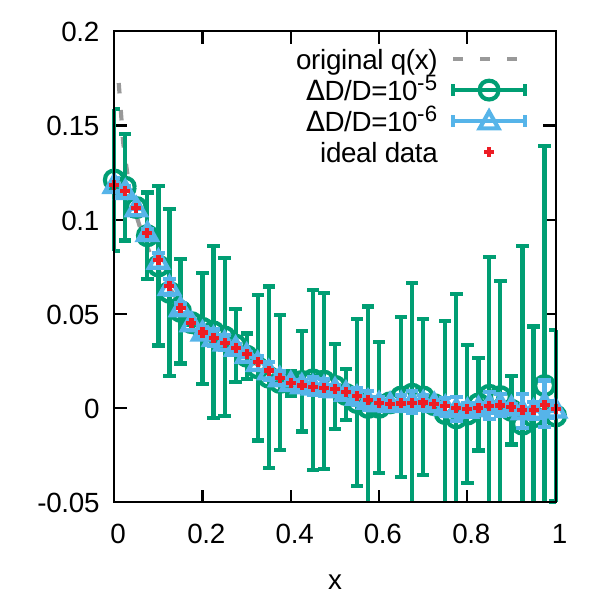}
		\caption{}
	\end{subfigure}
\caption{(a) The eigenvalues of the kernel matrix $\mathfrak{K}$ for different maximum values of available Ioffe time. Full access to the Brillouin zone corresponds to eigenvalues of order unity (orange open squares), while reduced access to limited Ioffe time values is accompanied by an exponential decrease in the eigenvalues, rendering inversion ill-conditioned. (b) Reconstruction (colored symbols) of a mock PDF (denoted $q(x)$, dashed line) from direct inversion in the presence of constant relative errors $\Delta D/D$. Note that as access to the Brillouin zone is reduced (left) full (center) $4/5\nu_{\rm max}$ and (right) $0.2 \nu_{\rm max}$ the reconstruction becomes unreliable even in the presence of only minute statistical errors (plots reprinted from \cite{Karpie:2019eiq}).}\label{fig:invprob}
\end{figure}

As was highlighted in Ref.~\cite{Karpie:2019eiq} when access to the full Brillouin zone is available, the eigenvalues of the matrix $\mathfrak{K}$ are all of the order unity and an inversion of the discretized \cref{eq:defpseudo} is well conditioned, i.e. small errors in the Ioffe time data are not amplified. This amounts to the orange open symbols in panel (a) of \cref{fig:invprob}. Once we reduce access to Ioffe times so that only a fraction of the Brillouin zone is covered, the eigenvalues of the kernel decay exponentially, indicating that an inversion will lead to exponential amplification of noise in the input data. And indeed as we see in panel (b) of \cref{fig:invprob}, the reconstruction (colored symbols) of a mock PDF $q(x)$ (dashed line) from direct inversion of the kernel matrix from noisy Ioffe-time data proceeds excellently when full access to the Brillouin zone is given (left).  It however becomes increasingly unreliable as access is reduced. Using a realistic estimate of coverage in Ioffe-time available in current lattice QCD simulations (right) shows that a naive inversion remain unreliable even if unrealistically small noise is present in the input data. For a recent detailed study of the challenges associated with the inverse Fourier transform see Ref.~\cite{xiong2025solving}.

Apparently the inverse problem we face presents a spectrum of difficulty. Let us therefore briefly survey what types of inverse problems exist and where our inverse problem falls. Using the language of probability we may define an inverse problem as the task of estimating the distribution of an unobserved property from observed data. The most benign form of such a broadly defined inverse problem is \textit{interpolation}, where we attempt to estimate unobserved quantities within the same function space of our measured data, in a region that is bounded by measured data. The presence of observed data places strong constraints on the possible values the interpolated quantities may take on. The next more challenging task would be that of \textit{extrapolation}, where we remain within the same function space but now attempt to estimate the distribution of unobserved quantities in a region which is not bounded by measured data. The ultimate challenge of an inverse problem arrives in the form of \textit{inference} where we are attempting to determine the distribution of unobserved data in a function space that is different from that in which observations are made. The necessary transformation between the function spaces usually filters out information in the forward direction and thus limits the sensitivity of measured data to constraint the solution of the inverse problem, on top of the uncertainties presented by extrapolation.

An ill-posed inverse problem, as defined by Hadamard presents two key challenges: the solution is not unique and the solution is sensitive to uncertainty in the observed data. In case that we are dealing with the extraction of PDFs from noisy and discrete lattice QCD matrix elements we must admit that we are indeed facing such an ill-posed inverse problem, as long as we do not have access to the full Brillouin zone in Ioffe time.  

A meaningful reconstruction of PDFs in the context of the inverse problem thus requires us to regularize the ill-posed problem, using additional knowledge we possess about the system, so called prior information (for a more detailed discussion of the role of prior information see e.g. \cite{Rothkopf:2018jaj}). The use of prior information allows us to select the most probable result from an otherwise degenerate set of multiple solutions. 

In turn we are faced with two sources of uncertainty in the reconstruction procedure. We must deal with the error inherent in the observed data (both statistical error from Monte-Carlo estimates and systematic errors from the renormalization procedure) and the uncertainty associated with the choice of a particular regularization.

It is here that there exist opportunities for fruitful collaboration between the communities focussing on the extraction of PDFs and those working on the determination of spectral functions of mesons at finite temperature. In both cases an ill-posed unfolding problem akin to \cref{eq:defpseudo} must be solved. The $T>0$ community brings to the table experience in the uncertainty quantification of inverse problems honed e.g. in the study of the in-medium spectra of heavy quarkonium states, where the identification of relevant features in the reconstructed spectrum is crucial to answer the core question of the field: up to which temperature does a vacuum bound state survive? The need to distinguish regularization artifacts (e.g. ringing) from physical changes in the spectral function led to a detailed study and careful treatment of the inverse problem (see e.g. Refs.~\cite{aarts2014bottomonium,Kim:2018yhk}). It became clear that both improvements in the input data quality, as well as a better understanding of regularization effects are key to arrive at a reliable physics interpretation. 

It is therefore encouraging to see on the side of the $T=0$ PDF community that there is continued progress in the design of improved matrix elements to be used as input for the reconstruction (see e.g. \cite{zhao2024transverse}) and the data quality is increasing steadily, by now allowing first extractions of challenging quantities, such as the gluon PDF (see Ref.~\cite{Good:2025daz}). In the context of regularization effects there is an ongoing discussion about different strategies of how to treat the limited reach in Ioffe-time where both parametric extrapolations and non-parametric priors are explored (see Refs.~\cite{dutrieux2025inverse,chen2025lamet,dutrieux2025comment}).

The appearance of various inverse problems in the study of hadron structure as well as nuclear matter under extreme conditions has lead to increased community interest in the topic, reflected in a series of workshops at the \href{https://websites.umass.edu/acfi/workshop/qcd-real-time-dynamics-and-inverse-problems/}{\texttt{Amherst Center}} , \href{https://indico.ectstar.eu/event/101/timetable/#20210913}{\texttt{ECT*}}, \href{https://indico.cern.ch/event/1313552/overview}{\texttt{CERN}} and most recently two workshops at \href{https://www.int.washington.edu/programs-and-workshops/24-88w}{\texttt{INT}} and \href{https://indico.mitp.uni-mainz.de/event/356/}{\texttt{MITP}}. The last two workshops explicitly highlight the question of uncertainty quantification.

\section{Tackling the inverse problem}

In the preceding section we identified the presence of an ill-posed inverse problem in the extraction of PDFs from lattice QCD. It is related to the fact that sparse data $\mathfrak Q$ with a finite error $\epsilon$ can be reproduced by a infinitely many PDFs $f(x)$ (non-uniqueness) and that the solution is sensitive to errors in input data (ill-conditioned). Over the past decades various approaches have been developed to tackle unfolding problems, such as the reconstruction of PDFs and spectral functions, all of which have in common that one needs to introduce regularization, i.e. prior information, to obtain a unique solution.

The simplest form of regularization is to assume a parametrized model for the function of interest. In the context of PDFs one often finds use of a phenomenologically inspired ansatz $f(x) = c x^a (1-x)^b$, where the prescribed functional form reduces the problem to finding three parameters from a standard $\chi^2$ fit. 

\subsection{Survey of non-parametric reconstruction methods}

Proceeding towards non-parametric reconstruction approaches, let us first consider linear methods, such as the Tikhonov regularization, the Backus-Gilbert (BG) method (see e.g. \cite{DelDebbio:2024sfa}) or Gaussian processes (GP) (see e.g. \cite{Horak:2021syv}). The BG method can be understood as performing the reconstruction in a limited function space given by the image of the Kernel $\mathfrak{K}$. Gaussian processes assume that observed and unobserved quantities are jointly Gaussian distributed according to a common GP prior (for a recent application see \cite{Medrano:2025cmg}). Importantly, linear methods only have very limited ability to incorporate prior information, such as key properties of various PDFs, such as positivity.

Fully non-linear approaches to unfolding have been developed based on Bayesian inference. In the context of in-medium spectral functions the Maximum Entropy Method (MEM) \cite{asakawa2001maximum}, as well as the Bayesian Reconstruction (BR) method \cite{burnier2013novel} have seen wide application over the past decade. Both methods introduce a regularization that is derived from a set of axioms. While the MEM was developed originally for two-dimensional image reconstruction in astronomy, the BR method uses axioms specifically tailored to the one-dimensional reconstruction problem at hand. And while the MEM axioms focus on avoiding the introduction of correlations in the reconstructed data where there are none in the input data, the BR method puts the smoothness of the reconstructed function center stage.

Recently interest sparked in non-Bayesian approaches, such as Nevanlinna \cite{Fei_2021} and Prony \cite{Huang:2022qsb} reconstruction, which on the one hand offer to incorporate prior  information about the analytic structure of the matrix elements used and (at least the latter) also involve a regularization mechanism related to singular value decomposition.

Last but not least let me also mention the reconstruction based on neural networks (NN). The use of machine learning for the reconstruction of PDFs and correspondingly for spectral functions was first proposed in Ref.~\cite{Karpie:2019eiq} and has since been explored in various studies (see e.g. Ref.~\cite{Aarts:2025gyp}).

Let us take a closer look at some of the reconstruction methods found in the literature on PDF and spectral function reconstruction.

\subsubsection{Backus-Gilbert method}

Given a convolution $Q(\nu)=\int dx K(x,\nu)q(x)$ we as asked to find linear combination of the data $Q$ to assemble the function $q$. Let us define the auxiliary quantity $A(\nu') = \int d\nu a(\nu',\nu) Q(\nu)$. After expressing the data in terms of the convolution $A$ is related to the sought after function $q$ 
\begin{align}
A(\nu') = \int dx \int d\nu a(\nu',\nu) K(x,\nu) q(x) = \int dx D(x-\nu') q(x)
\end{align}
Here the resolution function $D$ appears. Our task is to find linear combination matrix $a(\nu',\nu)$, which provides the best possible resolution, i.e. making $D(x-\nu')$ as peaked as possible. In case that $D$ can be turned into a genuine delta function our auxiliary quantity $A$ represents $q$. Of course the resolution achievable by this approach depends on the form of the kernel matrix $K$ and the quality of the data. 

Due to the functional form of the kernels appearing in the reconstruction of PDF and spectral functions the resolution function is often quite broad, meaning that the BG method is unable to resolve well localized peak structures. Such artificial smoothing is a known systematic artifact and needs to be accounted for in uncertainty quantification 

\subsubsection{Bayesian inference approaches}

Bayesian inference relies on the application of Bayes theorem
\begin{align}
P[f|Q,I] \propto P[Q|f,I] P[f|I]; \quad P[Q|f,I] = e^{-L}, \; L= \frac{1}{2}\sum_i ( Q(\nu_i)-Q(\nu_i)^f)^2/\sigma_i^2; \quad P[f|I]=e^S,\; S=S[f(x),m(x),\alpha(x)].
\end{align}
which states that the \textit{posterior} probability $P[q|Q,I]$ of some test function $q(x)$, to be the correct PDF, given input data $Q$ and prior information $I$, is obtained from a product of two terms, the likelihood probability $P[Q|q,I]$ and the prior probability $P[q|I]$. The likelihood encodes how probable it is that the observed data $Q$ was actually produced from a PDF of the form $q$. In case of lattice QCD simulations, where the matrix elements are estimated from subaverages of observables our input data is approximately Gaussian distributed and thus the likelihood is nothing but the standard $\chi^2$ likelihood. 

The prior probability on the other hand encodes how compatible our test function $q$ is with prior information. It performs the function of the regulator by penalizing functions $q$ that are at odds with key properties, such as positivity. The prior is conventionally parametrized using two parameters, the default model $m(x)$ and the uncertainty hyperparameter $\alpha(x)$. The former encodes where the prior probability takes on its maximum and $\alpha$ is a measure of the uncertainty in this value. Different Bayesian approaches to PDF reconstruction differ in the choice of prior probability, encoding different pieces of prior information. While the MEM uses the Shannon-Jaynes entropy as regulator $S^{\rm SJ}$, the BR method proposes a regulator functional $S^{\rm BR}$ that produces a prior probability in the form of the Gamma distribution (for a detailed discussion of different regulators see e.g. \cite{Rothkopf:2022ctl}). It is well known that in the presence of a small number of input datapoints the MEM in its state-of-the-art implementation according to Bryan introduces an artificial smoothening onto the reconstructed PDF (for a detailed discussion see \cite{Rothkopf:2011ef,Rothkopf:2020qqt}). The artifacts introduced by the BR method are very different. Its regulator is relatively weak compared to the MEM and thus may allow for artificial ringing to contaminate the reconstruction. Therefore considering both MEM and BR method results allows one to identify more reliably which features in a reconstructed spectral function correspond to physics encoded in the matrix elements.

Once the likelihood and regulator have been chosen according to input data and available prior information, one may either sample the posterior using modern Hybrid Monte-Carlo techniques or settle for a simple maximum a posteriori estimate of the most PDF $f_{\rm Bayes}$ by numerically optimizing the posterior $\delta P[f|Q,I] /\delta f= 0$. In the absence of data, by construction, the most PDF is given by the default model.

The benefit of the Bayesian analysis is that it puts both sources of uncertainty in the open. Statistical uncertainty in the input data may be assessed by resampling analysis, such as the statistical Jackknife. On the other hand, the dependence on the regularization is clear in the choice of the functional form and the parameters of the prior probability, all of which should be varied to assess the reliability of the reconstructed PDF during uncertainty quantification.

\subsubsection{Neural network reconstruction}

\begin{figure}
		\centering
		\includegraphics[scale=0.4]{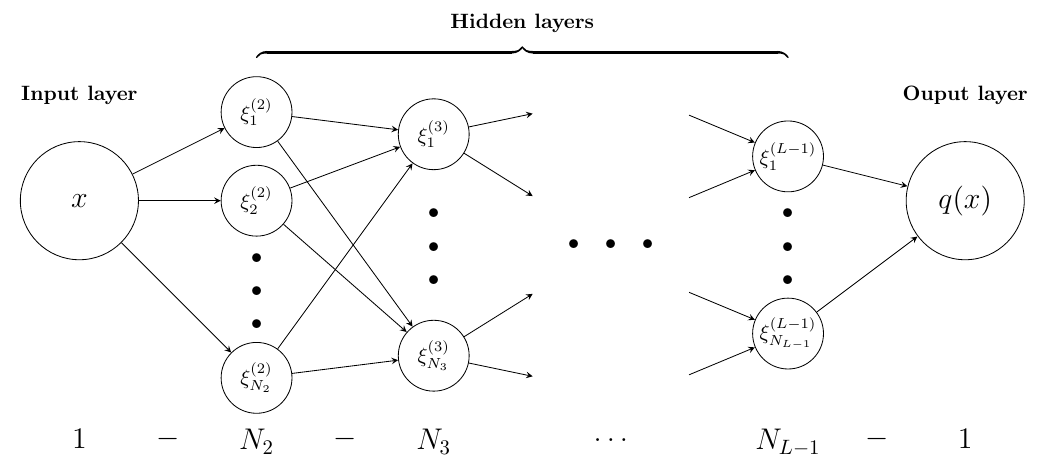}
		\caption{A possible application of neural networks in the reconstruction of PDFs: starting from a single parameter input layer representing the Bjorken-x value at which we wish to evaluate the PDF, a set of hidden layers is spanned which eventually are projected into a single output layer representing the PDF value (figure reprinted from \cite{Karpie:2019eiq}).}
\label{fig:MLPDF}
\end{figure}

A neural network can be used as a very expressive choice of basis functions for the reconstruction of a PDF, as proposed in Ref.~\cite{Karpie:2019eiq}. We wish to model the PDF by predicting its value at a given Bjorken-x parameter. To this end a single parameter input layer, representing said x-value, branches into a set of hidden layers, which ultimately are projected onto a single parameter output layer as sketched in \cref{fig:MLPDF}. The task at hand is to find the parameters $\xi_i^{(j)}$ of the hidden layers, given input data. 

In this form the NN represents a parameterization of the PDF, which contains more parameters than input data, therefore allowing for infinitely many possible parameter sets that will reproduce the input data. Thus we also need to specify some form of regularization, which is provided in the form of a training functional. This training functional will contain a part that favors reproducing the input data and some regularization term, often Tikhonov-like regularization, that favors small weights. In that sense, the reconstructions based on NN in the literature can be understood as a form of Bayesian inference with the prior probability (the regulator) encoded in the specific form of the training functional. Elucidating how the final result depends on the choice of training functional and training data is key to establish the full uncertainty budget. 

\subsection{Benchmarking PDF reconstruction approaches} 

\begin{figure}
		\centering
		\includegraphics[scale=0.4]{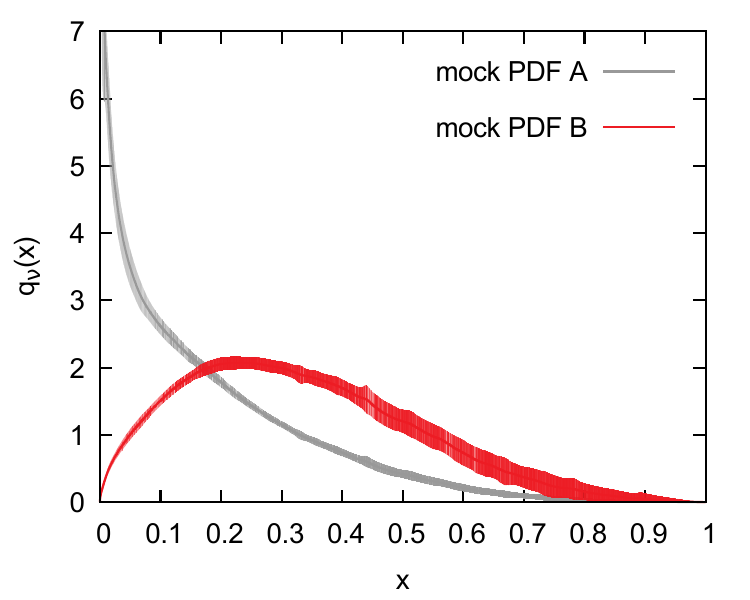}
		\includegraphics[scale=0.4]{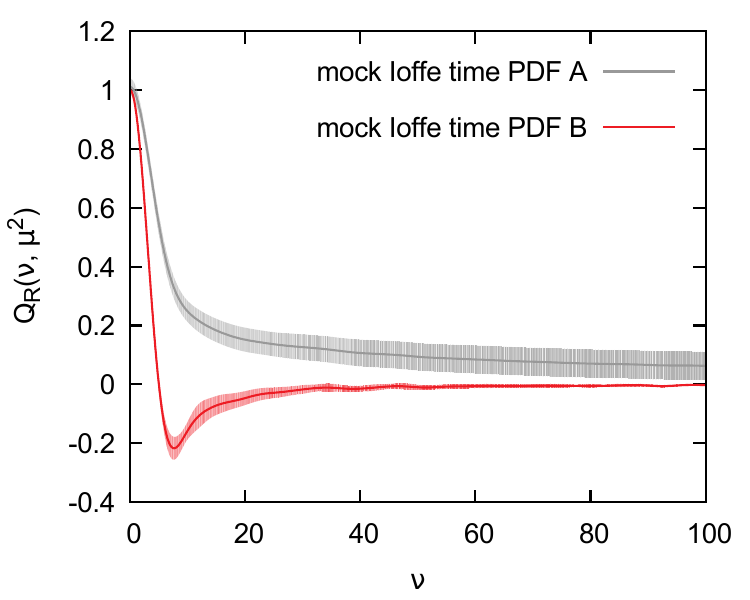}
		\caption{(left) Mock PDFs used in our benchmark study. Featuring similar convex behavior at values $x>0.4$, model PDF B (red) exhibits a downward trend leading to a vanishing intercept, while model PDF A (gray) exhibits a monotonous increase towards smaller Bjorken x. (right) The values of the matrix elements according to the two mock PDFs, evaluated over a large range of Ioffe time (plots reprinted from \cite{Karpie:2019eiq}).}
\label{fig:mockPDF}
\end{figure}

Having surveyed various reconstruction approaches, I will focus in the following on three methods that have been deployed in multiple studies in the past, the Backus-Gilbert reconstruction, as well as the MEM and BR method. Let us benchmark their performance in PDF reconstruction, based on mock PDFs which represent two scenarios discussed in the literature. As shown in the left panel of \cref{fig:mockPDF} both model PDFs similarly increase in value, as one decreases $x$. The key difference occurs around $x\approx0.4$, where model PDF B (red) begins to fall to intercept with the y-axis at the origin. This behavior is expected from phenomenology. The other model PDF A (gray) continues to rise, intercepting the y-axis at a finite value. The right panel of \cref{fig:mockPDF} shows the corresponding Ioffe-time data according to these two mock PDF models, plotted over a relatively large range of Ioffe time. 

For a realistic assessment of the performance of different reconstruction approaches we consider access to Ioffe-time matrix elements up to a maximum value of $\nu=10$, which is realistic given today's state-of-the-art lattices.

\begin{figure}
		\centering
		\includegraphics[scale=0.4]{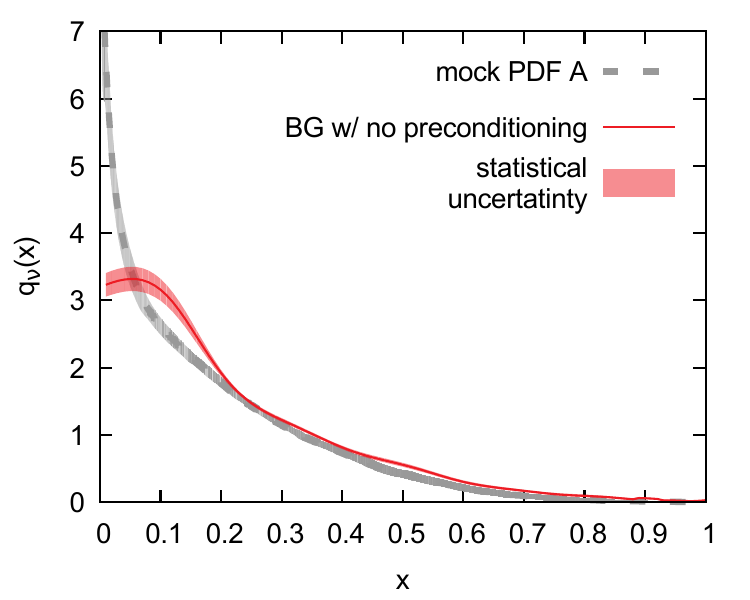}
		\includegraphics[scale=0.4]{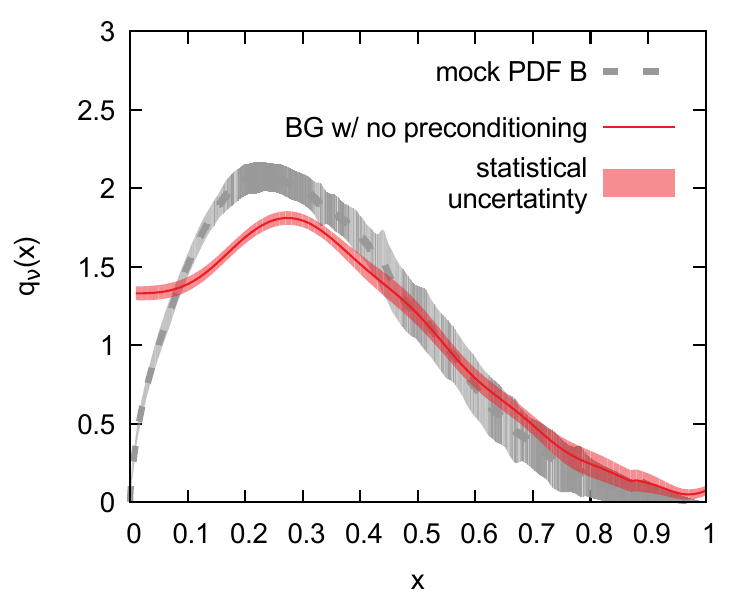}\\
		\includegraphics[scale=0.4]{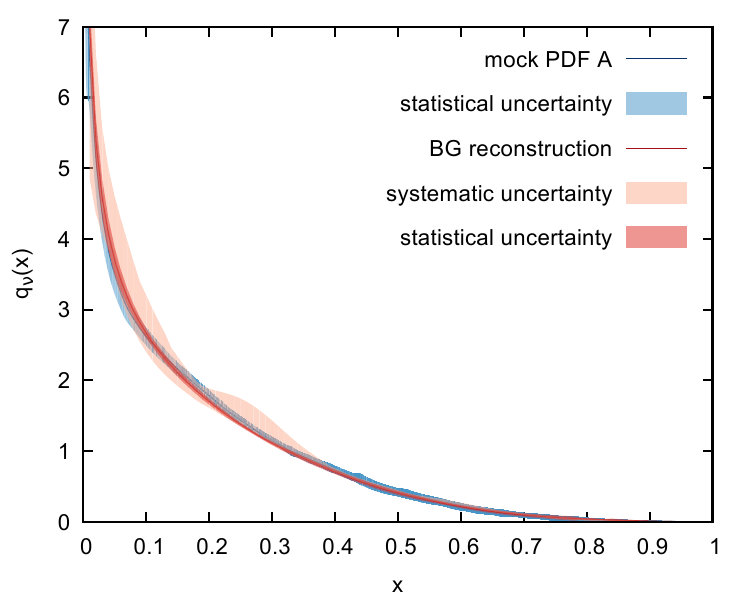}
		\includegraphics[scale=0.4]{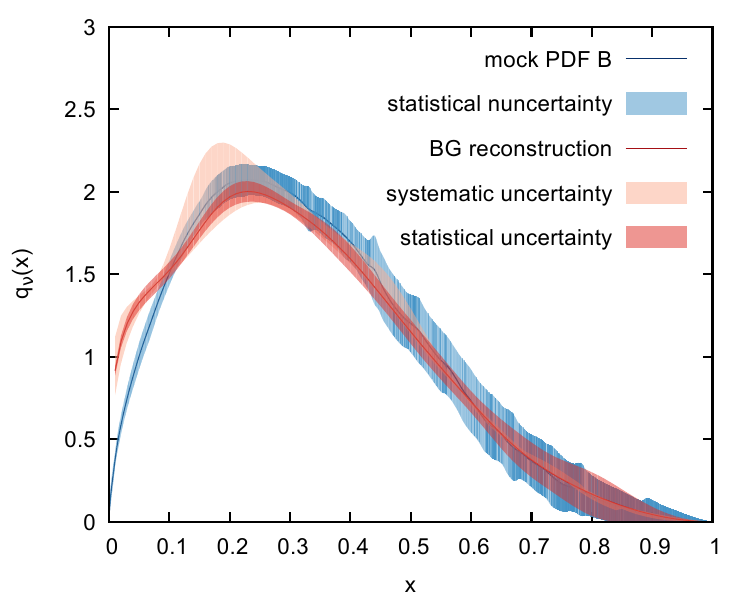}
		\caption{(top row) Backus-Gilbert reconstruction (red) of the mock PDFs (gray dashed) based on raw input data. Note the deviations of the reconstruction for $x<0.5$ from the true solution not captured by uncertainty bands. (bottom row) Improved BG reconstruction based on preconditioned data using fit data (plots reprinted from \cite{Karpie:2019eiq}).}
\label{fig:mockPDFBG}
\end{figure}

Deploying the Backus-Gilbert method on the raw data does not produce satisfactory results as seen in the top row of \cref{fig:mockPDFBG}. One reason for this result is that fact that the BG method is not required by construction to reproduce the input data. We find, as expected, that the method is unable to capture the peaked features of the mock PDFs, neither at small $x$ for mock PDF A (left), nor at intermediate $x$ for mock PDF B (right). To improve performance, we precondition the data via a model fit. I.e. we first fit the Ioffe-time data using the phenomenological model $f_{\rm fit}(x) = c x^a (1-x)^b$ and only ask the BG method to identify deviations from that fit by rescaling the kernel as $K=cos(\nu x) f_{\rm fit}(x)$. In turn the reconstruction is performed on the function $h(x)=q(x)/f_{\rm fit}(x)$, which is closer to a flat function. With preconditioning the BG method shows improved performance in the bottom row of \cref{fig:mockPDFBG}, even though it still struggles with the small-x region in the phenomenologically motivated mock PDF. Note that even with preconditioning the error estimate in the small x region does not manage to capture the true solution.

\begin{figure}
		\centering
		\includegraphics[scale=0.4]{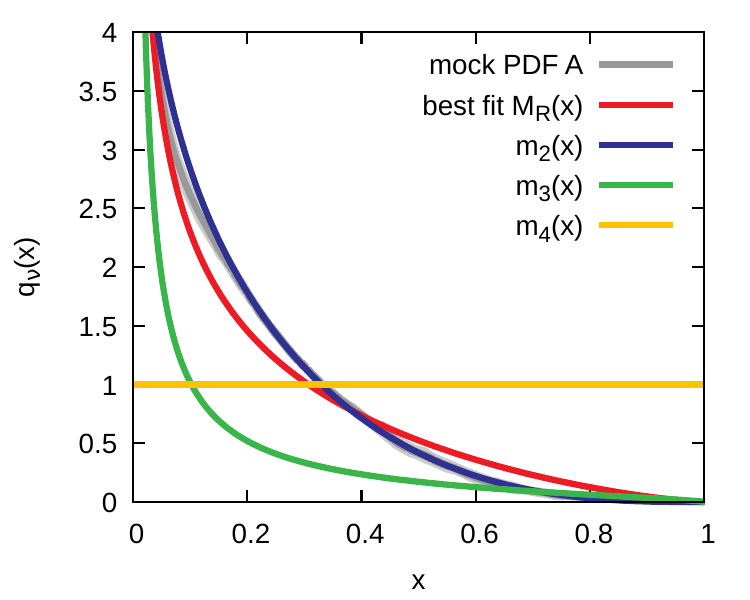}
		\includegraphics[scale=0.4]{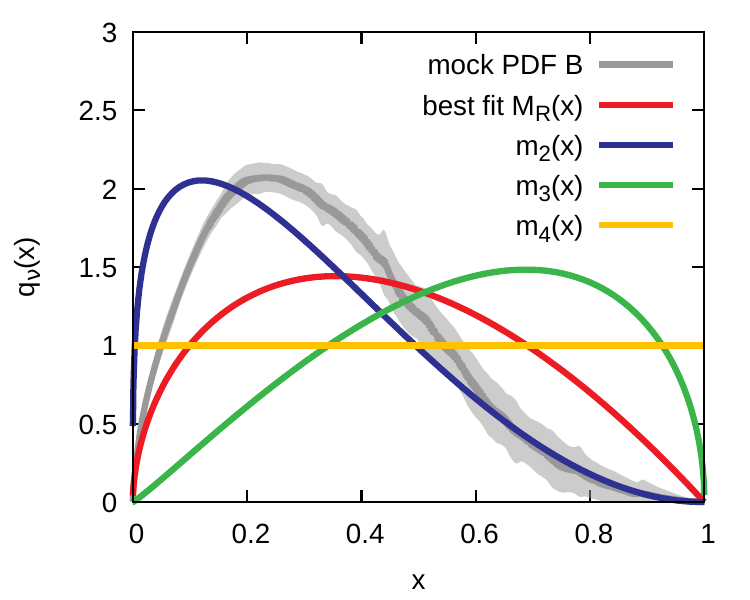}
		\caption{ Best phenomenological fit (red) obtained from using the function $f_{\rm fit}(x) = c x^a (1-x)^b$ to reproduce the Ioffe-time data of mock PDF A (left) and mock PDF B (right). For the data used in the fit see the right panel of \cref{fig:mockPDF}. In addition to the best fit result we also plot several deformations used as variations of the default model $m(x)$ in the Bayesian approaches to PDF reconstruction (plots reprinted from \cite{Karpie:2019eiq}).}
\label{fig:mockPDFfit}
\end{figure}

Let us now turn to the Bayesian approaches. Here we must provide a default model, the most probable estimate of the PDF a priori. This is achieved by performing the phenomenological model fit with $f_{\rm fit}(x) = c x^a (1-x)^b$, which is shown as red curve in \cref{fig:mockPDFfit}. Key to the uncertainty quantification of the Bayesian approaches is that we can easily deform the default model and assess the dependence of the final result of the reconstruction on the choice of $m(x)$. Several different choices for the default model are shown as colored curves. The total error budget of the Bayesian PDF reconstruction will consist of the Jackknife resampling error combined with the variance arising from using the different default models.

\begin{figure}
		\centering
		\includegraphics[scale=0.4]{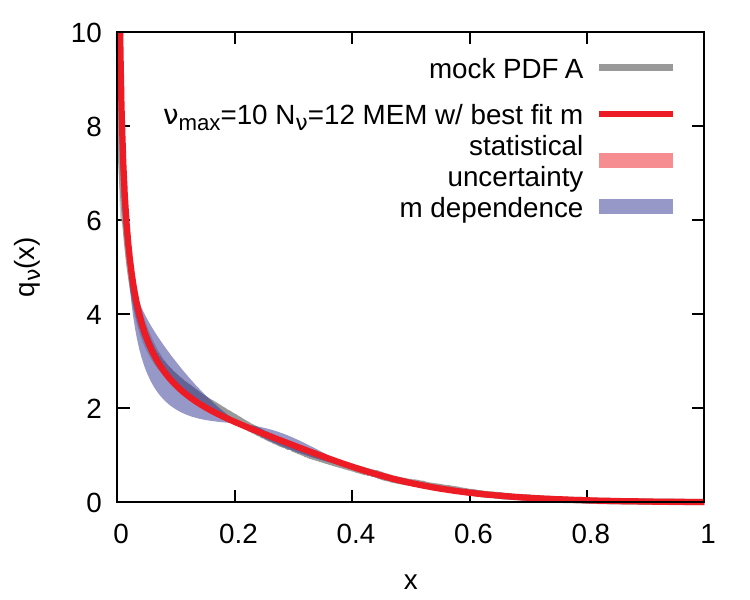}
		\includegraphics[scale=0.4]{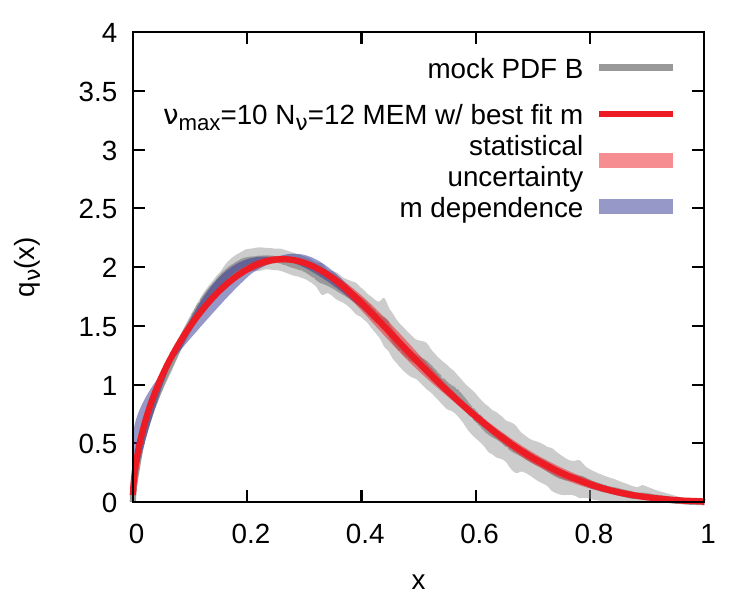}\\
		\includegraphics[scale=0.4]{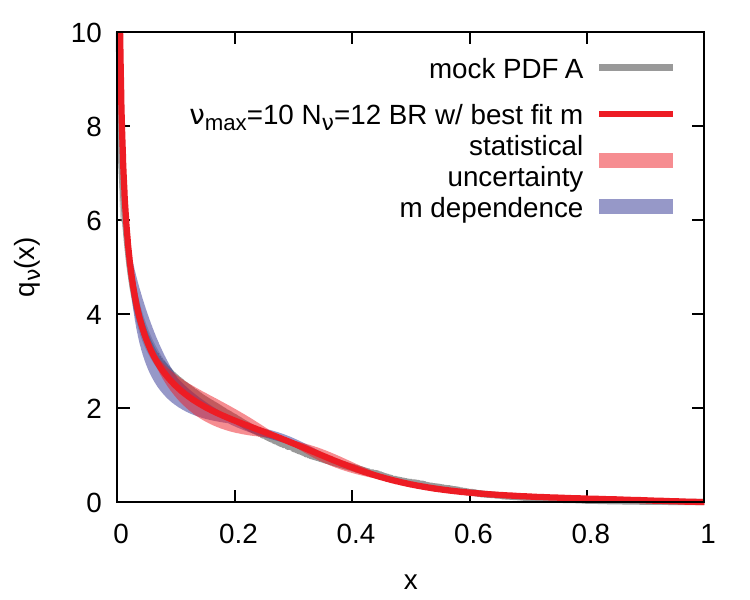}
		\includegraphics[scale=0.4]{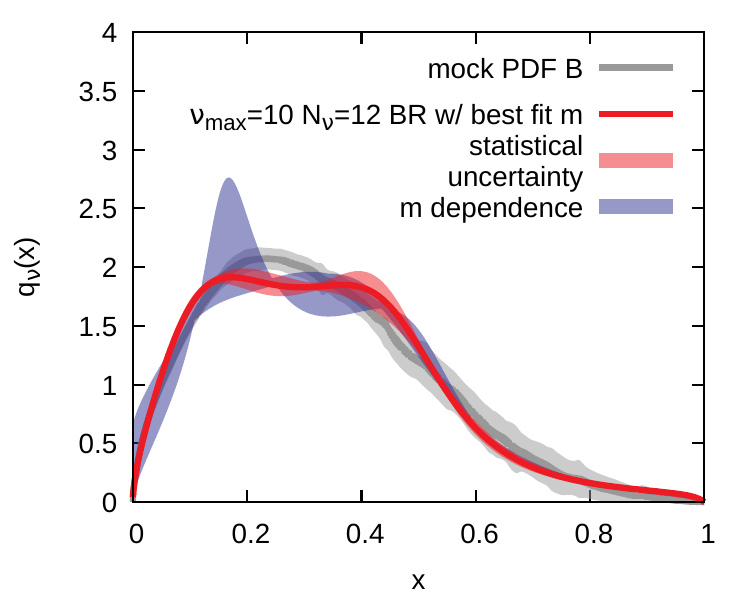}
		\caption{(top row) Reconstruction of the mock PDFs (gray dashed) using the Maximum Entropy Method. We find excellent agreement with the true PDF even though only a small number of datapoints and a limited range of Ioffe time is available. The smoothing property of the MEM here is beneficial to avoid artificial ringing. (bottom row) Reconstruction of the same data using the BR method. While in case of mock PDF A (left) the reconstruction is also excellent, the reconstruction of mock PDF B (right) suffers from some ringing artifacts (plots reprinted from \cite{Karpie:2019eiq}).}
\label{fig:mockPDFBR}
\end{figure}

As shown in the top row of \cref{fig:mockPDFBR}, for the small number of 12 datapoints used in the reconstruction, we find that the MEM provides excellent reproduction (red solid) of the underlying mock PDF (gray dashed). It performs very well both for model PDF A (left) and model PDF B (right). We understand that the artificial smoothing inherent in Bryan's MEM helps to stabilize the reconstruction result and to avoid ringing. 

The bottom row of \cref{fig:mockPDFBR} shows the reconstruction obtained based on the BR method. And while the reconstruction in case of model PDF A (left) is equally accurate as in case of the MEM, we find that for model PDF B (right) the BR method suffers from significant ringing artifacts. These ringing artifacts are associated with both the small reach in Ioffe time, as well as the small number of available datapoints. We have checked that the BR reconstruction improves significantly if one extends the available Ioffe time range to $\nu_{\rm max}=20$. 

\section{Conclusion}

Parton distribution functions encode vital information about the dynamical structure of hadrons. Accessing PDFs in lattice QCD is difficult due to the fact that lattice QCD simulations are carried out in Euclidean time, where the concept of lightcone is absent. Their extraction nevertheless has been shown to be possible in principle by approaching the lightcone from below, i.e. from the evaluation of space-like separated correlation functions at large momenta. Two competing implementations of this idea are currently explored in the form of quasi- and pseudo-PDFs. 

The inverse Fourier transform represents a key step in the analysis pipeline of both quasi and pseudo PDFs. Such a transformation is well-posed if the whole Brillouin zone is accessible. On the other hand if only a limited portion of the Brillouin zone is covered, the matrix representing the discretized IFT becomes ill conditioned rendering the determination of the PDF ill-posed. All approaches used to address an ill-posed inverse problem have in common that we must specify additional prior information to regularize the problem. Only then will we obtain a unique solution. 

There exist opportunities for fruitful collaboration between the communities focussed on the extraction of PDFs at $T=0$ and the community working on spectral function reconstruction for in-medium hadrons, as in both cases very similar ill-posed inverse problems arise. There is a long history in developing and applying statistical analysis methods for the purpose of spectral function reconstruction that can benefit the recent work on PDFs. In this proceeding, we surveyed the Backus-Gilbert method as an example of a linear method, the MEM and BR method as examples of general non-linear Bayesian methods and we briefly touched on the use of neural networks in the context of the reconstruction problem. Most reconstruction method can be formulated in the language of Bayesian inference, assuming different forms of prior information on the target function to be reconstructed. A key aspect of a successful reconstruction is not only to provide an estimate of the target PDF but also to determine the full uncertainty budget. Bayesian approaches are particularly suited to this end, as they make explicit the dependence of the solution on both the uncertainty of the data (likelihood) and the uncertainty in the choice of regularization (prior probability). 

We find that using realistic PDF models, converted into Ioffe time matrix elements for a mock reconstruction, the methods regularly used in the spectral function reconstruction community offer viable paths forward also in the reconstruction of PDFs. While the Backus Gilbert method as linear method remains limited in the prior information it can exploit, the BR and especially the MEM show good performance in the extraction of PDFs from limited Ioffe-time matrix elements.

The experience collected in the extraction of spectral functions from Euclidean correctors will thus serve a valuable role in the ongoing quest to extract PDFs from state-of-the-art lattice QCD ensembles.

A.R. galdly acknowledges support by Korea University through project K2503291 \textit{Ab-initio simulation of the real-time dynamics of non-relativistic fermions}, as well as projects K2511131 and K2510461

\bibliographystyle{elsarticle-num}
\bibliography{InverseProblemPDFLattice}

\end{document}